\documentclass[manuscript]{aastex}

\usepackage{natbib}
\usepackage{amsmath}

\begin{document}
\title{Modeling of differential rotation in rapidly rotating solar-type stars}
\author{H. Hotta and T. Yokoyama}
\affil{Department of Earth and Planetary Science, University of Tokyo,
7-3-1 Hongo, Bunkyo-ku, Tokyo 113-0033, Japan}
\email{ hotta.h@eps.s.u-tokyo.ac.jp}
\begin{abstract}
We investigate differential rotation in rapidly rotating solar-type
 stars by 
 means of an axisymmetric mean field model that was previously applied to
 the sun. This allows us to calculate the latitudinal entropy gradient
 with a reasonable physical basis.
 Our conclusions are as follows: (1) Differential rotation approaches
the Taylor-Proudman state when stellar rotation is faster
than solar rotation.
(2) Entropy gradient generated by the attached subadiabatic layer
 beneath the convection zone becomes
 relatively small with a large stellar angular velocity.
(3) Turbulent viscosity and turbulent angular momentum
 transport determine the spatial difference of angular velocity
 $\Delta\Omega$. (4) The results of our mean field model can explain
 observations of stellar differential rotation.
\end{abstract}
\keywords{Sun: interior --- Sun: rotation --- Stars: interior}
\section{INTRODUCTION}
Our sun has an eleven-year magnetic activity cycle, which is thought to be
sustained by the dynamo motion of internal ionized plasma,
i.e., a transformation of kinetic energy to magnetic energy
\citep{1955ApJ...122..293P}. Our understanding of the solar dynamo has
significantly improved during the past fifty years, and some kinematic
studies can now reproduce solar
magnetic features such as equatorward migration of sunspots and poleward
migration of the magnetic field
\citep{1995A&A...303L..29C,1999ApJ...518..508D,2001A&A...374..301K,2005LRSP....2....2C,2010ApJ...709.1009H,2010ApJ...714L.308H}. 
The most important mechanism of the solar dynamo is the $\Omega$
effect, the bending of pre-existing poloidal magnetic fields by
differential rotation and the generation of toroidal magnetic fields.
Thus, the distribution of the differential rotation in the convection zone is
a significant factor for the solar dynamo.
Using helioseismology, it has recently been shown that the solar internal
differential rotation is in a non-Taylor-Proudman state
\citep[see review by][]{2003ARA&A..41..599T}, meaning
the iso-rotation surfaces are {\it not} parallel to the axis.
\par
Based on solar observations,
it is known that Ca H-K fluxes can be a signature of stellar chromospheric
activity, and such chromospheric signatures are in correlation
with magnetic activity.
\cite{1968ApJ...153..221W,1978ApJ...226..379W} and
\cite{1995ApJ...438..269B} discuss a class of
stars that
shows a periodic variation in Ca H-K
fluxes, which suggests that they have a magnetic cycle similar to our sun.
It is natural to conjecture that such magnetic activity is maintained
by dynamo action.
Various studies have been conducted to investigate the relationship
between stellar
angular velocity $\Omega_0$ and its latitudinal difference 
$\Delta\Omega$ i.e., $\Delta\Omega\propto\Omega_0^n$, where the
suggested range
of $n$ is 
$0 <n < 1$
\citep{1996ApJ...466..384D,2003A&A...398..647R,2005MNRAS.357L...1B}.
This means that the angular velocity difference
$\Delta\Omega$ increases and the
relative difference $\Delta\Omega/\Omega_0$ decreases with increases in
the stellar
rotation rate $\Omega_0$.\par
In this paper, we investigate differential rotation in rapidly
rotating stars using a mean field framework. Our study is based on the
work of \cite{2005ApJ...622.1320R}, in which he suggests the importance
of the role of the subadiabatic layer below the convection zone in order
to maintain a non-Taylor-Proudman state in the Sun.
The aim of this paper is to use a mean field  model to analyze firstly
the dependence of the
morphology of differential
rotation on stellar angular velocity, and secondly the physical
process which
determines the observable angular velocity difference $\Delta \Omega$.
According to our knowledge, this is the first work which
systematically discusses the application of Rempel's (2005b) solar
model to stars.
\par
Other research adopts another approach to the use of mean field models
for the analysis of differential rotation in
rapidly-rotating stars \citep{1995A&A...299..446K,2001A&A...366..668K}.
In these studies, the non-Taylor-Proudman state is sustained by
anisotropy of turbulent thermal conduction. This anisotropy is generated by
the effects of stellar rotation on convective turbulence.\par
Three-dimensional numerical studies on stellar differential
rotation also exist
\citep{2008ApJ...689.1354B,2009AnRFM..41..317M}.
In these studies, they resolve stellar thermal driven convection and can
calculate a self-consistent turbulent angular momentum transport and
anisotropy of turbulent thermal conductivity.
The subadiabatic layer below the convection zone, however, is not included.
The effects of anisotropy of turbulent
thermal conductivity and the subadiabatic layer are discussed in this paper.

\section{MODEL}
Using numerical settings similar to those of Rempel's (2005b),
we solve the axisymmetric hydrodynamic equations in spherical geometry
$(r,\theta)$, where $r$ is the radius, and $\theta$ is the colatitude.
The basic assumptions are as follows.
\begin{enumerate}
 \item A mean field approximation is adopted. All processes on the
       convective scale
       are parameterized. Thus, the coefficients for turbulent viscosity,
       turbulent heat conductivity, and turbulent angular momentum
       transport are explicitly given in the equations.
 \item The perturbations of the density and pressure associated with
       differential rotation are small,
       i.e., $\rho_1\ll \rho_0$ and $p_1\ll p_0$. Here $\rho_0$ and $p_0$
       denote the reference state density and pressure respectively,
       whereas $\rho_1$
       and $p_1$ are the perturbations. We neglect the second-order
       terms of these quantities.
       Note that the perturbation of
       angular velocity ($\Omega_1$) and meridional flow ($v_r$,
       $v_\theta$) are not small.
 \item Since the reference state is assumed to be in an energy flux balance,
       the entropy equation includes only perturbations. 
\end{enumerate} 
\subsection{Equations}
We do not use the anelastic approximation here.
The equations in an inertial frame can be expressed as
\begin{eqnarray}
&& \frac{\partial \rho_1}{\partial t}=
-\frac{1}{r^2}\frac{\partial }{\partial r}(r^2v_r\rho_0)
-\frac{1}{r\sin \theta}\frac{\partial }{\partial \theta}(\sin\theta
v_\theta \rho_0)\label{continuity},\\
&& \frac{\partial v_r}{\partial t}=
-v_r\frac{\partial v_r}{\partial r}
-\frac{v_\theta}{r}\frac{\partial v_r}{\partial \theta}
+\frac{v_\theta^2}{r}
-\frac{1}{\rho_0}
\left[
\rho_1 g+\frac{\partial p_1}{\partial r}
\right]
+(2\Omega_0\Omega_1+\Omega_1^2)r\sin^2\theta+\frac{F_r}{\rho_0}\label{vx},\\
&& \frac{\partial v_\theta}{\partial t}=
-v_r\frac{\partial v_\theta}{\partial r}
-\frac{v_\theta}{r}\frac{\partial v_\theta}{\partial \theta}
-\frac{v_rv_\theta}{r}
-\frac{1}{\rho_0}\frac{1}{r}\frac{\partial p_1}{\partial \theta}
+(2\Omega_0\Omega_1+\Omega_1^2)r\sin\theta\cos\theta+\frac{F_\theta}{\rho_0}\label{vy},\\
&& \frac{\partial \Omega_1}{\partial t}=
-\frac{v_r}{r^2}\frac{\partial}{\partial r}[r^2(\Omega_0+\Omega_1)]
-\frac{v_\theta}{r\sin^2\theta}\frac{\partial }{\partial \theta}
[\sin^2\theta(\Omega_0+\Omega_1)]
+\frac{F_\phi}{\rho_0 r\sin\theta}\label{om1},\\
&& \frac{\partial s_1}{\partial t}=
-v_r\frac{\partial s_1}{\partial r}
-\frac{v_\theta}{r}\frac{\partial s_1}{\partial \theta}
+v_r\frac{\gamma\delta}{H_p}
+\frac{\gamma-1}{p_0}Q
+\frac{1}{\rho_0 T_0}\mathrm{div}(\kappa_\mathrm{t}\rho_0T_0\mathrm{grad}
s_1)\label{se1},
\end{eqnarray}
where  $\Omega_0$ is a constant value that represents the angular
velocity of the rigidly rotating
radiative zone.
 We set it as a parameter in Table
\ref{param}. $\gamma$ is the ratio of specific heats, with the value
for an ideal gas being $\gamma=5/3$. $\kappa_\mathrm{t}$ is the
coefficient of turbulent thermal conductivity.
 $\delta=\nabla-\nabla_\mathrm{ad}$
represents superadiabaticity, where $\nabla=d(\ln T)/d(\ln p)$
(see \S \ref{back}).
 $g$ denotes gravitational acceleration.
Following from this, the perturbation of pressure $p_1$ and
pressure scale height $H_p$ are expressed as
\begin{eqnarray}
&& p_1=p_0
\left(
\gamma\frac{\rho_1}{\rho_0}+s_1
\right), \\
&& H_p=\frac{p_0}{\rho_0 g}.
\end{eqnarray}
$s_1$ is dimensionless entropy normalized by the specific heat
capacity at constant volume $c_\mathrm{v}$.
Turbulent viscous force ${\bf F}$ follows from
\begin{eqnarray}
 F_r=\frac{1}{r^2}\frac{\partial }{\partial r}(r^2R_{rr})
  +\frac{1}{r\sin\theta}\frac{\partial }{\partial \theta}(\sin\theta
  R_{\theta r})
  -\frac{R_{\theta\theta}+R_{\phi\phi}}{r},
\end{eqnarray}
\begin{eqnarray}
 F_\theta=\frac{1}{r^2}\frac{\partial }{\partial r}(r^2R_{r\theta})
  +\frac{1}{r\sin\theta}\frac{\partial }{\partial \theta}(\sin\theta
  R_{\theta \theta})
  +\frac{R_{r\theta}-R_{\phi\phi}\cot\theta}{r},
\end{eqnarray}
\begin{eqnarray}
 F_\phi=\frac{1}{r^2}\frac{\partial }{\partial r}(r^2R_{r\phi})
  +\frac{1}{r\sin\theta}\frac{\partial }{\partial \theta}(\sin\theta
  R_{\theta \phi})
  +\frac{R_{r\phi}+R_{\theta\phi}\cot\theta}{r},
\end{eqnarray}
with the Reynolds stress tensor
\begin{eqnarray}
 R_{ik}=\rho_0
\left[
\nu_\mathrm{tv}
\left(
E_{ik}-\frac{2}{3}\delta_{ik}\mathrm{div}{\bf v}
\right)
+\nu_\mathrm{tl}\Lambda_{ik}
\right].\label{reynolds}
\end{eqnarray}
Here $\nu_\mathrm{tv}$ is the coefficient of turbulent viscosity and
$\nu_\mathrm{tl}$ is the coefficient of the $\Lambda$ effect
\citep{1995A&A...299..446K},  a non-diffusive angular
momentum transport caused by turbulence.
$\nu_\mathrm{tv}$ and $\nu_\mathrm{tl}$
are expected to have the same value, since both effects are caused by
turbulence, i.e., thermal driven convection. We discuss this in more detail
in \S \ref{diffusivity}.
$E_{ik}$ denotes the deformation tensor, which is given in spherical
coordinates by 
\begin{eqnarray}
 &&E_{rr}=2\frac{\partial v_r}{\partial r},\\
 &&E_{\theta\theta}=2\frac{1}{r}\frac{\partial v_\theta}{\partial
  \theta}+2\frac{v_r}{r},\\
 &&E_{\phi\phi}=\frac{2}{r}(v_r+v_\theta\cot\theta), \\
 &&E_{r\theta}=E_{\theta r}=r\frac{\partial }{\partial r}
\left(\frac{v_\theta}{r}\right)+\frac{1}{r}\frac{\partial v_r}{\partial
\theta},\\
 &&E_{r\phi}=E_{\phi r}=r\sin\theta\frac{\partial \Omega_1}{\partial
  r},\\
 &&E_{\theta\phi}=E_{\phi\theta}=\sin\theta\frac{\partial
  \Omega_1}{\partial \theta}.
\end{eqnarray}
An expression for the $\Lambda$ effect ($\Lambda_{ik}$) is given later.
The amount of energy that is converted by the Reynolds stress from
kinematic energy to internal energy is given by
\begin{eqnarray}
 Q=\sum_{i,k}\frac{1}{2}E_{ik}R_{ik}.
\end{eqnarray}
\subsection{Background Stratification}\label{back}
\begin{figure}[htbp]
 \epsscale{1.}
 \plotone{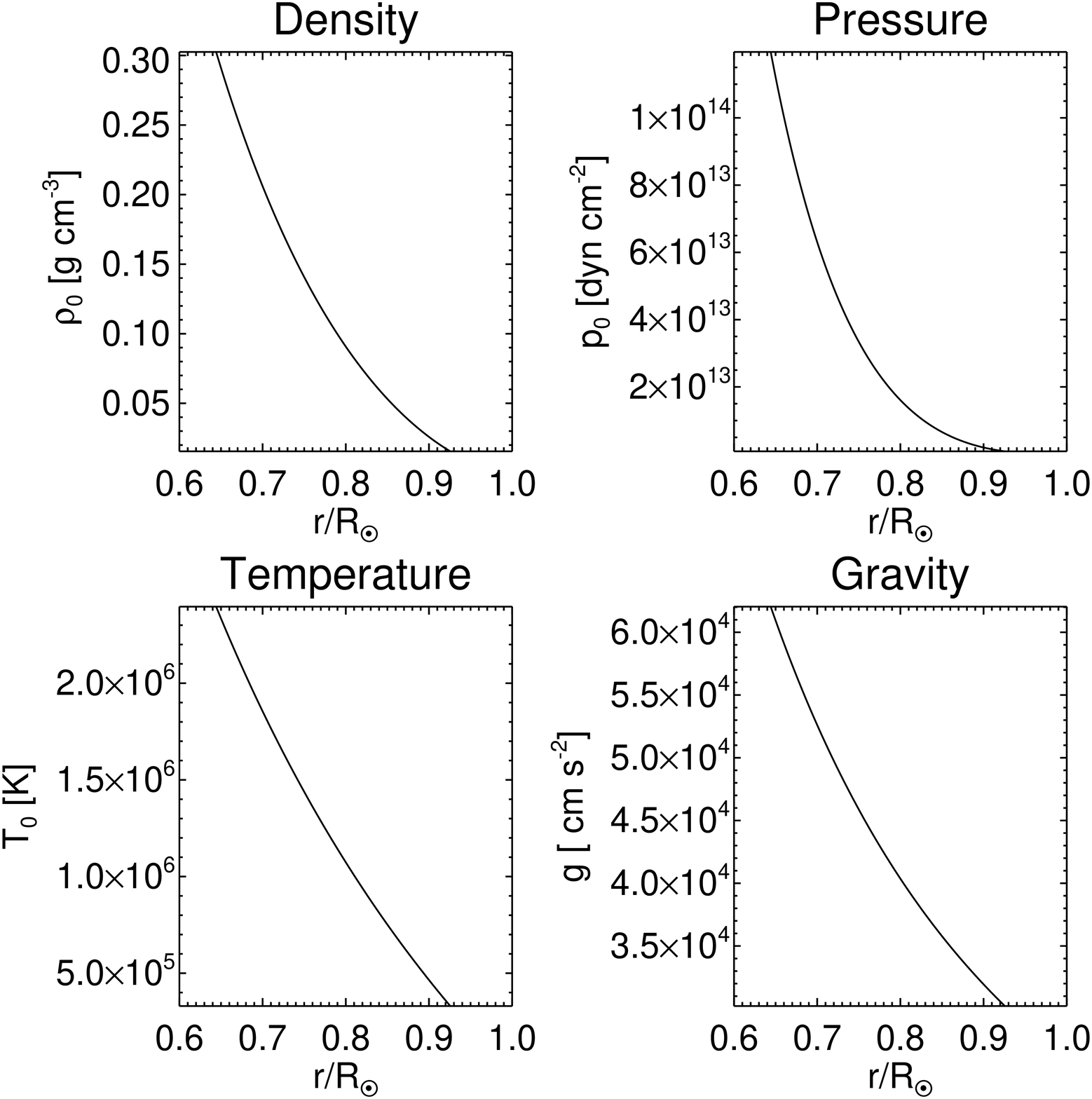}
 \caption{Profiles of density, pressure and temperature
 as a function of radial distance in the reference state. This
 stratification is adiabatic.\label{background}}
\end{figure}

We use an adiabatic hydrostatic stratification for the spherically
symmetric reference state of $\rho_0$, $p_0$ and $T_0$. 
Gravitational acceleration is assumed to have $\sim r^{-2}$
dependence, since the radiative zone ($r<0.65R_\odot$) has most of the solar
mass. 
This is expressed as,
\begin{eqnarray}
&& \rho_0(r)=\rho_\mathrm{bc}
\left[
 1+\frac{\gamma-1}{\gamma}\frac{r_\mathrm{bc}}{H_\mathrm{bc}}
 \left(\frac{r_\mathrm{bc}}{r}-1\right)
\right]^{1/(\gamma-1)},\\
&& p_0(r)=p_\mathrm{bc}
\left[
 1+\frac{\gamma-1}{\gamma}\frac{r_\mathrm{bc}}{H_\mathrm{bc}}
 \left(\frac{r_\mathrm{bc}}{r}-1\right)
\right]^{\gamma/(\gamma-1)},\\
&& T_0(r)=T_\mathrm{bc}
\left[
 1+\frac{\gamma-1}{\gamma}\frac{r_\mathrm{bc}}{H_\mathrm{bc}}
 \left(\frac{r_\mathrm{bc}}{r}-1\right)
\right],\\
&& g(r)=g_\mathrm{bc}
\left(
 \frac{r}{r_\mathrm{bc}}
\right)^{-2},
\end{eqnarray}
where $\rho_\mathrm{bc}$, $p_\mathrm{bc}$, $T_\mathrm{bc}$,
$H_\mathrm{bc}=p_\mathrm{bc}/(\rho_\mathrm{bc}g_\mathrm{bc})$
and $g_\mathrm{bc}$ denote the
values at the base of the convection zone $r=r_\mathrm{bc}$ of density,
pressure, temperature, pressure scale height and
gravitational acceleration, respectively.
In this study we use $r_\mathrm{bc}=0.71R_\odot$, with 
$R_\odot$ representing the solar radius ($R_\odot=7\times10^{10}\ \mathrm{cm}$).
We adopt solar values $\rho_\mathrm{bc}=0.2\ \mathrm{g\ cm^{-3}}$,
$p_\mathrm{bc}=6\times10^{13}\ \mathrm{dyn\ cm^{-2}}$,
$T_\mathrm{bc}=mp_\mathrm{bc}
/(k_\mathrm{B}\rho_\mathrm{bc})\sim1.82\times10^6\
\mathrm{K}$ and $g_\mathrm{bc}=5.2\times10^4\ \mathrm{cm\ s^{-2}}$,
where $k_\mathrm{B}$ is the Boltzmann constant, and $m$ is the mean
particle mass. Fig. \ref{background} shows the profiles of background
density, pressure and temperature, and gravitational acceleration.\par
\begin{figure}[htbp]
 \epsscale{1.}
 \plotone{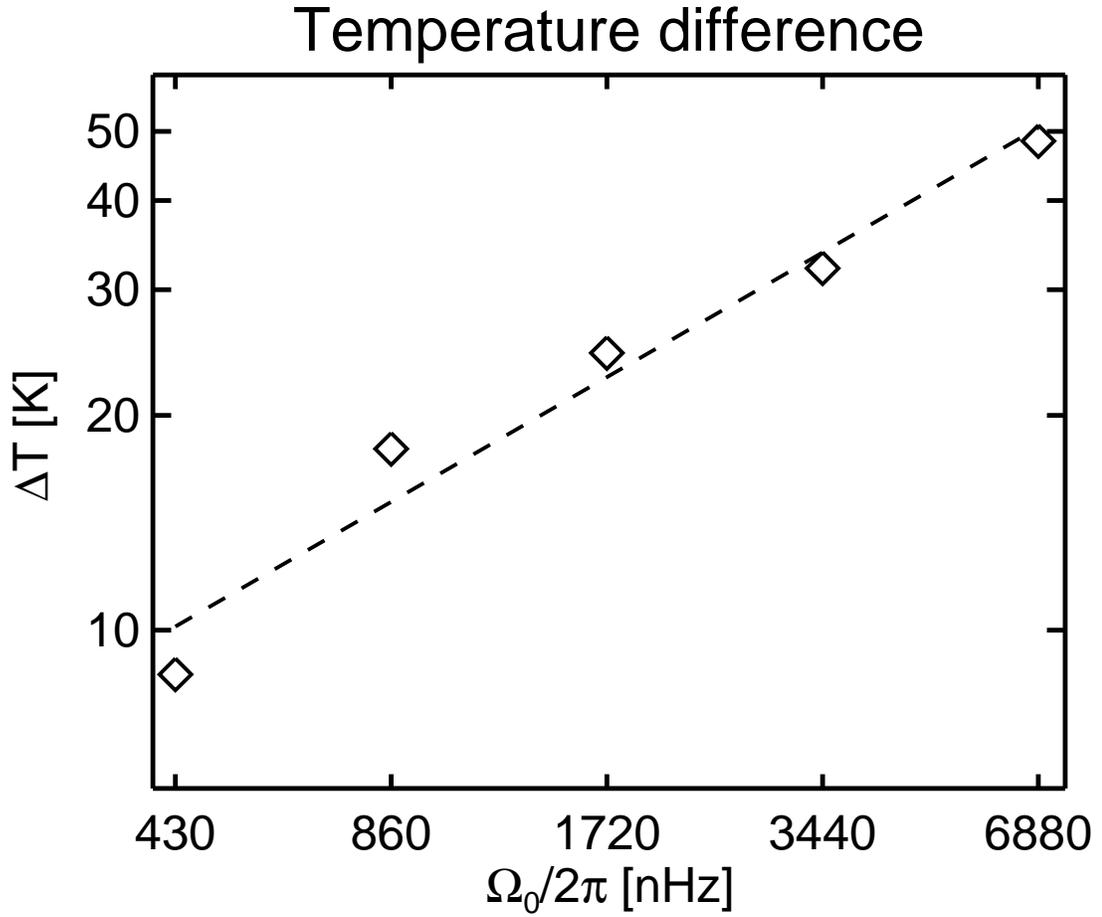}
 \caption{Temperature difference at the base of the convection zone
 ($r=0.71R_\odot$) as a function of stellar angular velocity
 ($\Omega_0/2\pi$). 
The dashed line is the fit to the results showing a power-law function
 with an index of 0.58.\label{entropy}}
\end{figure}

Although the real sun's
stratification is not adiabatic in the convection zone, our reference
state is valid, since
the absolute value of superadiabaticity is small.
In order to include the deviation from adiabatic stratification,
we assume superadiabaticity $\delta$
has the following profile:
\begin{eqnarray}
 \delta=\delta_\mathrm{conv}+\frac{1}{2}(\delta_\mathrm{os}-\delta_\mathrm{conv})
\left[
1-\tanh
\left(
\frac{r-r_\mathrm{tran}}{d_\mathrm{tran}}
\right)
\right].
\end{eqnarray}
Here $\delta_\mathrm{os}$ and $\delta_\mathrm{conv}$ denote the values of superadiabaticity in the
overshoot region. $r_\mathrm{tran}$ and $d_\mathrm{tran}$ denote the
position and the steepness of the transition toward the subadiabatically
stratified overshoot region, respectively.
Superadiabaticity in convection zone is define as
\begin{eqnarray}
 \delta_\mathrm{conv}=\delta_\mathrm{c}\frac{r-r_\mathrm{sub}}{r_\mathrm{max}-r_\mathrm{sub}},
\end{eqnarray}
where  $r_\mathrm{max}$ denotes the location of the upper boundary.
We specify
$\delta_\mathrm{os}=-1.5\times10^{-5}$, $r_\mathrm{tran}=0.725R_\odot$,
$r_\mathrm{sub}=0.8R_\odot$ and
$d_\mathrm{tran}=d_\mathrm{sub}=0.0125R_\odot$ in our simulations. 
$\delta_c$ is took as a free parameter.
The entropy gradient can be expressed as
\begin{eqnarray}
 \frac{ds_0}{dr}=-\frac{\gamma\delta}{H_p}.
\end{eqnarray}
The third term
of eq. (\ref{entropy}), $v_r\gamma\delta/H_p$, includes the effect of
deviations from
adiabatic stratification. The term indicates that an upflow (downflow) can make
negative (positive) entropy perturbations in the subadiabatically
stratified layers $(\delta<0)$. 

\subsection{Diffusivity Profile}\label{diffusivity}
\begin{figure}[htbp]
 \epsscale{1.}
 \plotone{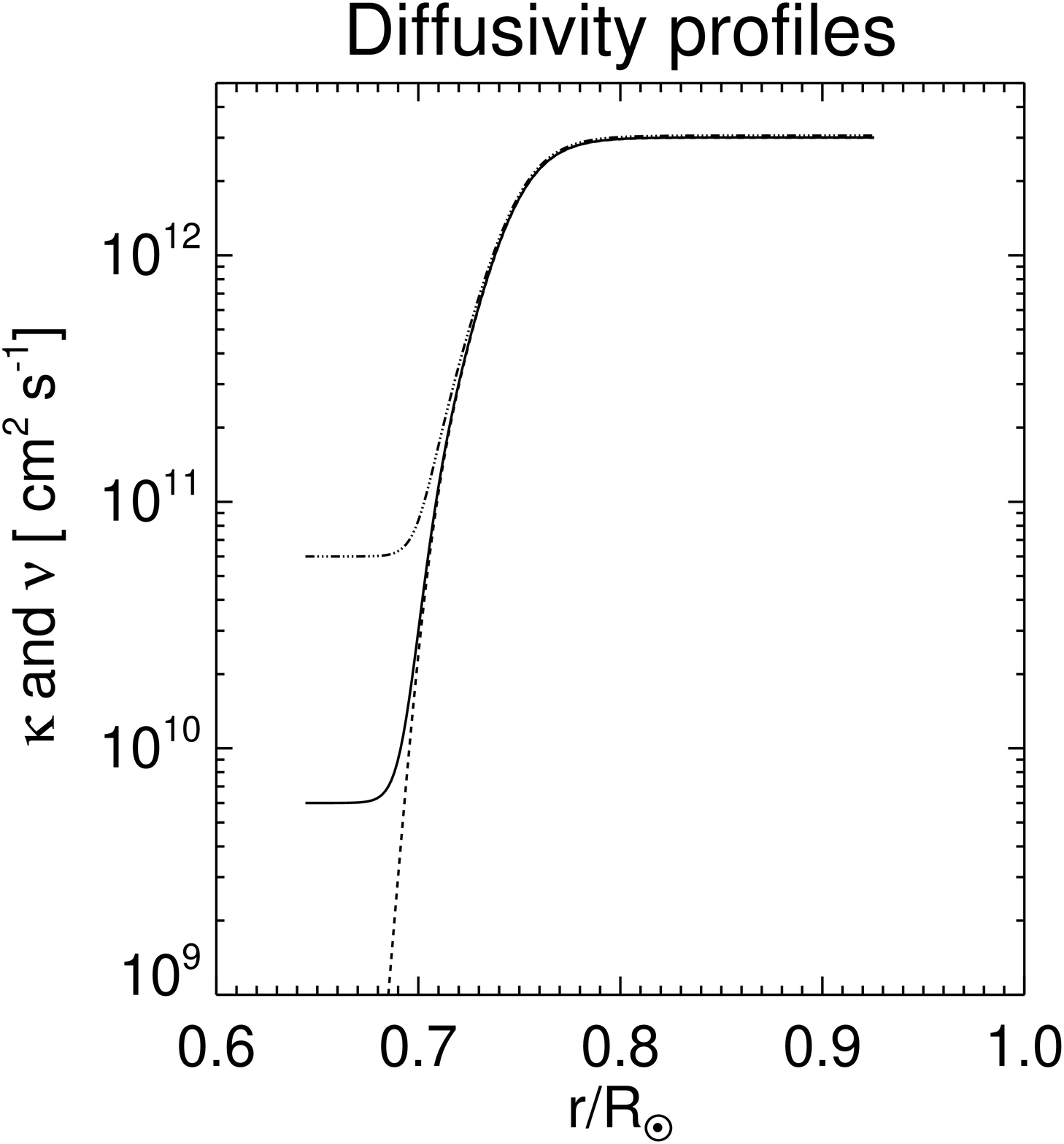}
 \caption{Profiles of diffusivity as a function of radial
 distance for cases 1-5 and 9-11. The solid line denotes the coefficient
 of turbulent
 conductivity $\kappa_\mathrm{t}$\label{diffusive}. The dashed line
 denotes the coefficient of the $\Lambda$ effect $\nu_\mathrm{tl}$. The
 dash and three dots line denotes the coefficient of turbulent
 viscosity $\nu_\mathrm{tv}$.
}
\end{figure}

We assume the coefficients of turbulent viscosity and thermal
conductivity to be constant within the convection zone, and these 
smoothly connect with the values of the overshoot region.
We assume that the diffusivities only depend on the radial coordinate:
\begin{eqnarray}
&& \nu_\mathrm{tv}=\nu_\mathrm{os}+\frac{\nu_\mathrm{0v}}{2}
\left[
1+\tanh
\left(
\frac{r-r_\mathrm{tran}+\Delta}{d_{\kappa\nu}}
\right)
\right]f_c(r),\\
&& \nu_\mathrm{tl}=\frac{\nu_\mathrm{0l}}{2}
\left[
1+\tanh
\left(
\frac{r-r_\mathrm{tran}+\Delta}{d_{\kappa\nu}}
\right)
\right]f_c(r),\\
&& \kappa_\mathrm{t}=\kappa_\mathrm{os}+\frac{\kappa_0}{2}
\left[
1+\tanh
\left(
\frac{r-r_\mathrm{tran}+\Delta}{d_{\kappa\nu}}
\right)
\right]f_c(r),
\end{eqnarray}
with
\begin{eqnarray}
&& f_c(r)=\frac{1}{2}
\left[
1+\tanh
\left(
\frac{r-r_\mathrm{bc}}{d_\mathrm{bc}}
\right)
\right],\\
&& \Delta=d_{\kappa\nu}\tanh^{-1}(2\alpha_{\kappa\nu}-1),
\end{eqnarray}
where $\nu_\mathrm{0v}$, $\nu_\mathrm{0l}$ and $\kappa_0$ are the values
of the turbulent diffusivities within the convection zone, and
$\nu_\mathrm{os}$ and $\kappa_\mathrm{os}$ are the values in the
overshoot region. We specify
$\nu_\mathrm{0l}=\kappa_\mathrm{0l}=3\times10^{12}\ \mathrm{cm^2\
s^{-1}}$,
$\nu_\mathrm{os}=6\times10^{10}\ \mathrm{cm^2\ s^{-1}}$ and
$\kappa_\mathrm{os}=6\times10^{9}\ \mathrm{cm^2\ s^{-1}}$, and we treat
$\nu_\mathrm{0v}$ as a parameter.
$\alpha_{\kappa\nu}$ specifies the values of the turbulent diffusivities
at $r=r_\mathrm{tran}$, i.e.,
$\nu_\mathrm{tv}=\nu_\mathrm{os}+\alpha_{\kappa\nu}\nu_\mathrm{0v}$,
$\nu_\mathrm{tl}=\alpha_{\kappa\nu}\nu_\mathrm{0l}$ and
$\kappa_\mathrm{t}=\kappa_\mathrm{os}+\alpha_{\kappa\nu}\kappa_\mathrm{0}$
at $r=r_\mathrm{tran}$. $d_\mathrm{bc}$ and $d_{\kappa\nu}$ are the widths
of transition. We specify $\alpha_{\kappa\nu}=0.1$,
$d_\mathrm{bc}=0.0125R_\odot$ and $d_{\kappa\nu}=0.025R_\odot$.
As already mentioned, the coefficients for
turbulent viscosity and the $\Lambda$ effect are different in our model
from
those of Rempel's (2005b).
There are two reasons for this. 
One is that we intend to investigate the influence of both effects
on stellar differential rotation separately (see \S
\ref{delta_omega}). The other reason is that
the formation of a tachocline in a reasonable amount of time requires a finite
value (though small) for the coefficient of
turbulent viscosity even in the radiative zone, in which there
is likely to be weak
turbulence \citep{2005ApJ...622.1320R}.
Fig. \ref{diffusive} shows the profiles of $\nu_\mathrm{tv}$,
$\nu_\mathrm{tl}$ and $\kappa_\mathrm{t}$.
\subsection{The $\Lambda$ Effect}\label{s:lambda}
In this study we adopt the non-diffusive part of the Reynolds stress, called
the $\Lambda$ effect. The $\Lambda$ effect transports angular momentum and
generates differential rotation. The $\Lambda$ effect tensors are
expressed as
\begin{eqnarray}
 && \Lambda_{r\phi}=\Lambda_{\phi
  r}=+L(r,\theta)\cos(\theta+\lambda), \\
&& \Lambda_{\theta\phi}=\Lambda_{\phi
  \theta}=-L(r,\theta)\sin(\theta+\lambda),
\end{eqnarray}
where $L(r,\theta)$ is the amplitude of the $\Lambda$ effect and
$\lambda$ is the
inclination of the flux vector with respect to the rotational axis.
We use for the amplitude of the $\Lambda$ effect the expressions
\begin{eqnarray}
&& f(r,\theta)=\sin^l\theta\cos\theta\tanh
\left(\frac{r_\mathrm{max}-r}{d}
\right),
\\
&&
 L(r,\theta)=\Lambda_0\Omega_0\frac{f(r,\theta)}{\mathrm{max}|f(r,\theta)|}
\label{lambda},\label{amp_lambda}
\end{eqnarray}
where 
$d=0.025R_\odot$. $\lambda$ and $\Lambda_0$  are  free-parameters.
The value of $l$ needs to be equal to or larger than 2 to ensure regularity near
the pole, so we set $l=2$.
The $\Lambda$ effect does not depend on $v_r$, $v_\theta$ or $\Omega_1$,
meaning it is a stationary effect. We emphasize that the $\Lambda$ effect
depends on stellar angular velocity $\Omega_0$, since the $\Lambda$ effect
is generated by turbulence and Coriolis
force. The more rapidly the star rotates, the more angular momentum the
$\Lambda$ effect can transport.
The dependence of $\Lambda_0$ and $\lambda$ on stellar angular velocity
is discussed in \S \ref{variation}.
\subsection{Numerical Settings}\label{numerical}
Using the modified Lax-Wendroff scheme with TVD artificial viscosity
\citep{davis1984tvd},
we solve Equations (\ref{continuity})-(\ref{se1}) numerically for the
northern hemisphere of the meridional plane in 
$0.65R_\odot < r <0.93R_\odot$ and $0 < \theta < \pi/2$.
 We use a
uniform resolution of $200$ points in the radial direction and
$400$ points in
the latitudinal direction in all of our simulations. 
Each simulation run is conducted until it reaches a stationary state.
All the variables $\rho_1$, $v_r$, $v_\theta$,
$\Omega_1$ and $s_1$ are equal to zero in the initial condition.
At the top boundary ($r=0.93R_\odot$) we
adopt stress-free boundary conditions for $v_r$, $v_\theta$ and
$\Omega_1$ and set the derivative of $s_1$ to zero:
\begin{eqnarray}
&& \frac{\partial v_r}{\partial r}=0,\\
&& \frac{\partial}{\partial r}
\left(
\frac{v_\theta}{r}
\right)=0,\\
&&\frac{\partial \Omega_1}{\partial r}=0,\\
&&\frac{\partial s_1}{\partial r}=0.
\end{eqnarray}
The boundary conditions for $v_r$, $v_\theta$ and $s_1$ at the lower boundary
($r=0.65R_\odot$) are the same as those at the top boundary.
Differential rotation connects with the rigidly rotating core at the lower
boundary, so we adopt $\Omega_1=0$ there. 
At both radial boundaries, we set $\rho_1$ to make the right side of
eq. (\ref{vx}) equal zero.
At the pole and the equator
($\theta=0$ and $\pi/2$) we use the symmetric boundary condition:
\begin{eqnarray}
&& \frac{\partial \rho_1}{\partial \theta}=0, \\
&& \frac{\partial \Omega_1}{\partial \theta}=0, \\
&& \frac{\partial v_r}{\partial \theta}=0, \\
&& v_\theta=0,\\
&& \frac{\partial s_1}{\partial \theta}=0.
\end{eqnarray}\par
Due to the low Mach number of the expected flows, a direct compressible
simulation is problematic, so adopting the same technique as
\cite{2005ApJ...622.1320R}, we reduce the speed of
sound by multiplying the right side of eq. (\ref{continuity}) by
$1/\zeta^2$.
The equation of continuity is therefore replaced with
\begin{eqnarray}
 \frac{\partial \rho_1}{\partial t}+
\frac{1}{\zeta^2}\mathrm{div}(\rho_0{\bf v})=0.
\end{eqnarray}
The speed of sound then becomes $\zeta$ times smaller than the original
speed. We use $\zeta=200$ in all our calculations. 
This technique can be used safely in our present study since
we only discuss stationary states, so the factor $\zeta$ becomes
unimportant.
The validity of this technique is carefully discussed by
\cite{2005ApJ...622.1320R}.
We test our code by reproducing the results presented by
\cite{2005ApJ...622.1320R} and check the numerical convergence by
runs with different grid spacings.
After checking and cleaning up at every time step,
conservation of total mass, total angular momentum and total
energy are maintained through the simulation runs.
\section{Stellar Differential Rotation and the Taylor-Proudman
  Theorem}\label{differential}
In this section, based on the work of \cite{2005ApJ...622.1320R}, we explain how the subadiabatically stratified region
can generate solar-like differential rotation. 
The $\phi$ component
of the vorticity equation can be expressed as
\begin{eqnarray}
 \frac{\partial \omega_\phi}{\partial t}=[...]
+r\sin\theta\frac{\partial \Omega^2}{\partial z}
-\frac{g}{\gamma r}\frac{\partial s_1}{\partial \theta},\label{therm01}
\end{eqnarray}
where $\Omega=\Omega_0+\Omega_1$, and the $z$ axis represents the
rotational axis.
The inertial term and the diffusion term are neglected.
If the last term of eq. (\ref{therm01}) is zero, meaning there is no variation
in entropy in the latitudinal direction, then $\partial
\Omega^2/\partial z =0$ in a stationary state, which is the Taylor-Proudman
state. Solar-like differential rotation is generated in four
stages.
\begin{enumerate}
 \item In the northern hemisphere, the $\Lambda$ effect transports
       angular momentum in the negative $z$
       direction and  generates a
       negative $\partial \Omega^2/\partial z$.
 \item The negative $\partial \Omega^2/\partial z$ generates a negative
       $\omega_\phi$ due to Coriolis force. This counter-clockwise
       meridional flow corresponds to a
       negative $v_r$ (downflow) at high latitudes and a positive
       $v_r$ (upflow) at low latitudes.
 \item As we mentioned in Section \ref{back}, 
       downflow (upflow) generates positive
       (negative) entropy perturbations in the subadiabatically stratified
       layer beneath the
       convection zone ($\delta<0$). Meridional flow can
       generate positive entropy perturbations at high latitudes
       and negative entropy perturbations at low latitudes.
       Therefore, $\partial s_1/\partial \theta$ becomes negative in the
       overshoot region.
 \item The negative $\partial s_1/\partial \theta$ also keeps $\partial
       \Omega^2/\partial z$ negative in a stationary state.
\end{enumerate}
The profile of angular velocity in the convection zone is determined
by a balance of angular momentum transport from meridional flow
and a reduction in meridional flow from buoyancy force at the
subadiabatic layer.
\section{RESULTS AND DISCUSSION}
We run simulations for seventeen cases, with Table \ref{param} showing the
parameters for each case.
\subsection{Stellar Differential Rotation}\label{taylor}
\begin{figure}
\begin{center}
\begin{tabular}{cc}
\begin{minipage}{0.4\hsize}
\begin{center}
 \plotone{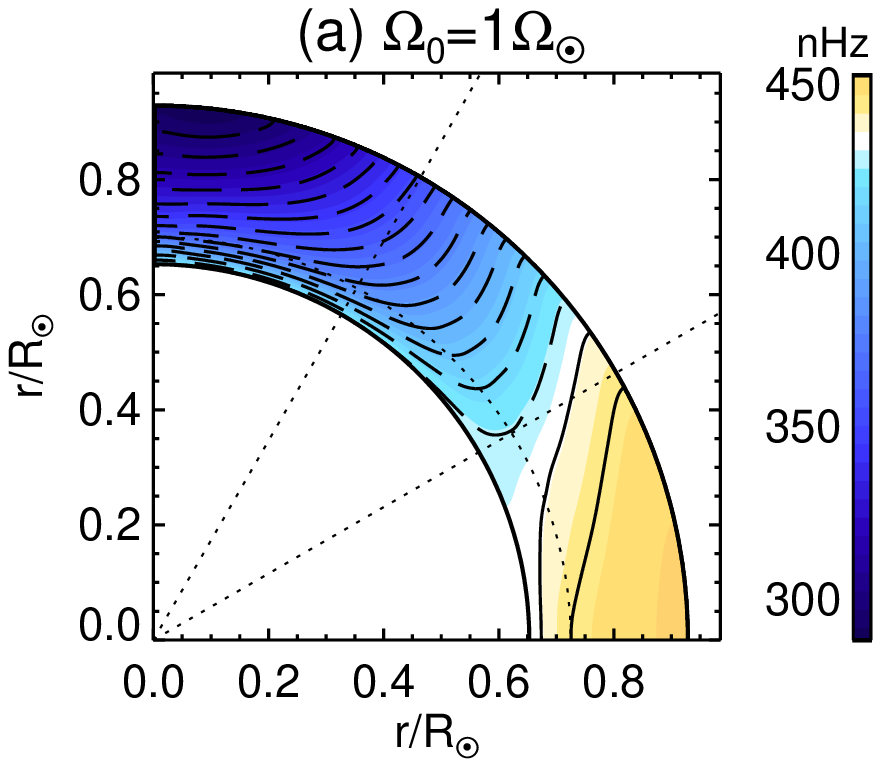}
\end{center}
\end{minipage}
\begin{minipage}{0.4\hsize}
\begin{center}
 \plotone{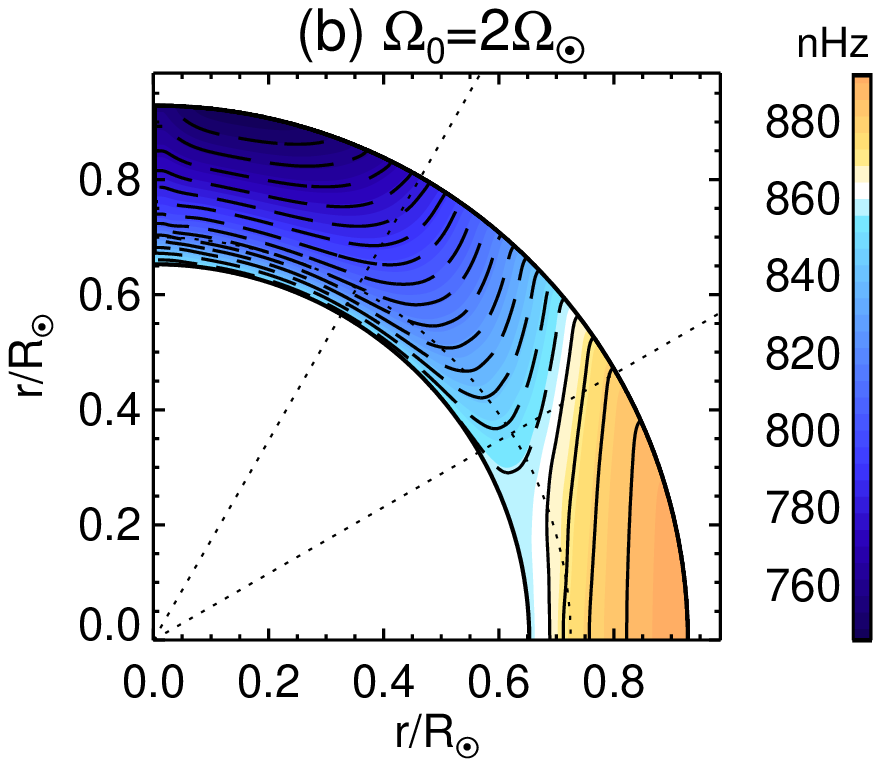} 
\end{center}
\end{minipage}
\end{tabular}
\\
\begin{tabular}{cc}
\begin{minipage}{0.4\hsize}
\begin{center}
 \plotone{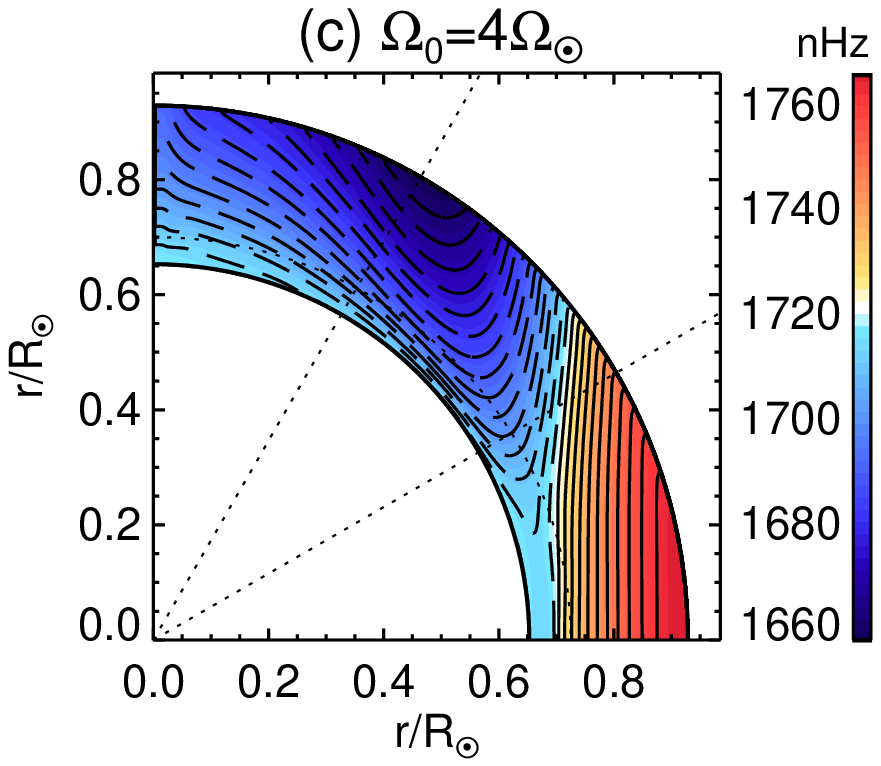}
\end{center}
\end{minipage}
\begin{minipage}{0.4\hsize}
\begin{center}
 \plotone{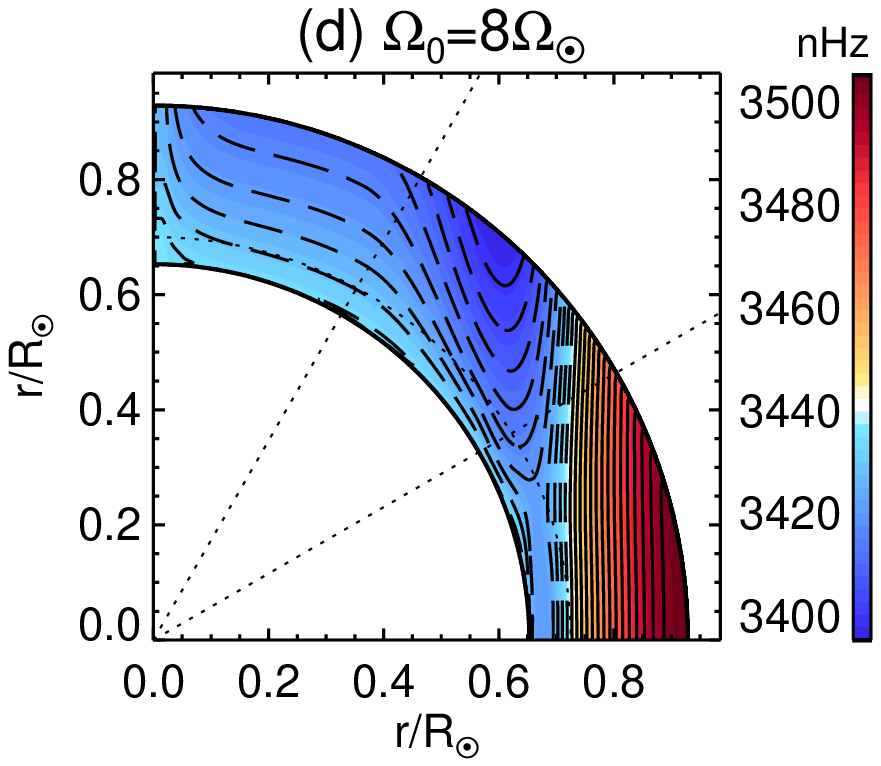} 
\end{center}
\end{minipage}
\end{tabular}

\begin{tabular}{c}
\begin{minipage}{0.4\hsize}
\begin{center}
 \plotone{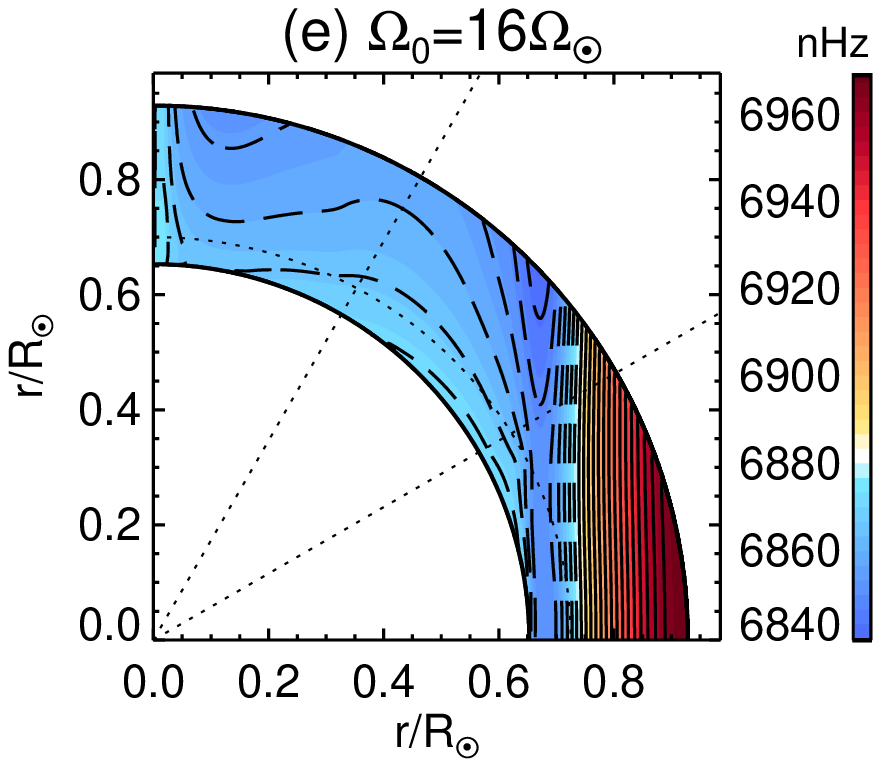}
\end{center}
\end{minipage}
\end{tabular}
\end{center}
\caption{
 Rotation profiles of the simulation results. Panels (a)-(e) correspond
 to  cases 1-5, respectively. The stellar rotation rate for each case is
 given at the top of each panel.
 The area of red and solid lines
 (blue and dashed lines) rotates faster (slower) than the rigidly
 rotating core at the bottom boundary.
 Color  bars are given for angular velocity
 $\Omega/2\pi=(\Omega_0+\Omega_1)/2\pi$ in the unit of nHz.
 The dotted lines in each panel indicate the
 base of the convection zone ($r=0.71R_\odot$) and the colatitudes
 $\theta=30^\circ$ and $\theta=60^\circ$.  
 \label{rapid}}
\end{figure}

\begin{figure}[htbp]
 \epsscale{1.}
 \plotone{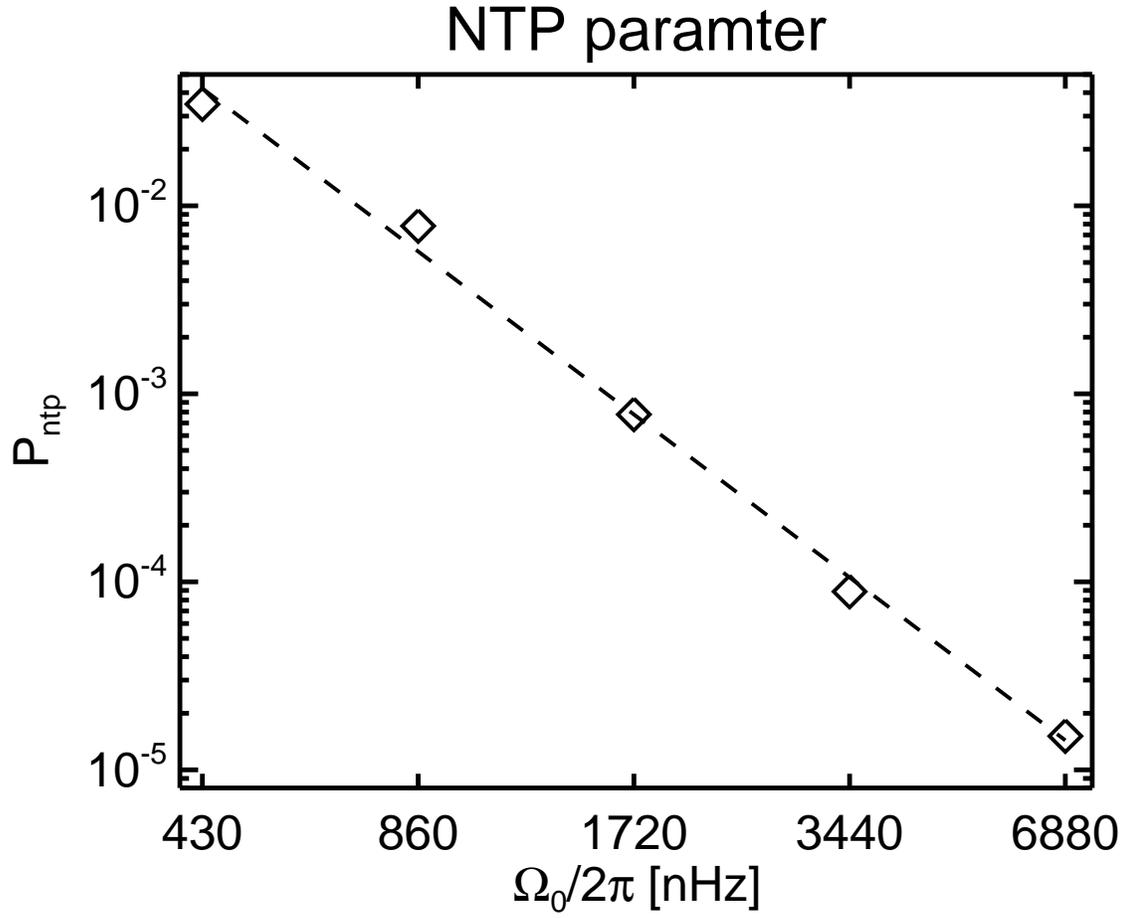}
 \caption{NTP parameter as a
 function of stellar angular velocity $\Omega_0/2\pi$.
 The dashed line is the fit to the results showing a power-law function
 with an index of $-2.9$.\label{npp}}

\end{figure}

In this section, we discuss the cases with angular velocities up to 16
times the solar value (represented by $\Omega_\odot$), placing an
emphasis on the morphology of
stellar differential rotation.
Fig. \ref{rapid} shows the results of our
calculations which correspond to cases 1-5 in Table \ref{param}. It is
found that the larger stellar angular velocity is, the more
likely it is for
differential rotation to be in the Taylor-Proudman state, in which the
contour lines of
the angular velocity are parallel to the rotational axis. 
To evaluate these results quantitatively, we define a
parameter which
denotes the morphology of differential rotation. We call it
the Non-Taylor-Proudman parameter (hereafter the NTP parameter), which is
expressed as
\begin{eqnarray}
 P_\mathrm{ntp}=\frac{1}{R_\odot^2\Omega_0^2}\int\frac{\partial
  \Omega_1^2}{\partial z}dV
  =\frac{1}{R_\odot^2\Omega_0^2}\int
\left(
\cos\theta\frac{\partial }{\partial r}
-\frac{\sin\theta}{r}\frac{\partial }{\partial \theta}
\right)
\Omega_1^2dV,
\end{eqnarray}
where $\Omega_0$ is the angular velocity of the radiative zone.
When the NTP parameter is zero, differential rotation is in the
Taylor-Proudman state. Conversely, differential rotation is far from the
Taylor-Proudman state with a large absolute value of the NTP parameter. The value
of the NTP parameter
with various stellar angular velocities is shown in
Fig. \ref{npp}. The
NTP monotonically decreases with increases in stellar angular
velocity. These results
indicate that with large stellar angular velocity values,
differential rotation approaches the Taylor-Proudman state.
These results are counter-intuitive, however, since we do not expect
differential rotation to approach the Taylor-Proudman state with
increasing stellar angular velocity values, since the
$\Lambda$ effect, which is a driver of the deviation from the Taylor-Proudman
state, is proportional to stellar angular velocity $\Omega_0$.
These are the most significant findings of this paper, so
hereafter in this section we discuss these unexpected results.
\par
We next discuss the temperature difference between the equator and the
pole at the base of the convection zone ($r=0.71R_\odot$). 
Since temperature is given as a function of entropy by
\begin{eqnarray}
 T_1=\frac{T_0}{\gamma}
\left[s_1+(\gamma-1)\frac{p_1}{p_0}
\right],
\end{eqnarray}
and it is easier to measure than entropy, we use it here for discussing
the thermal
structure of the simulation results in the convection zone.
Further, although it is mentioned in \S \ref{differential} that entropy
gradient is crucial for breaking the Taylor-Proudman constraint,
the temperature difference can be used as its proxy. Fig. \ref{entropy} shows the
relationship between stellar angular velocity $\Omega_0$ and
temperature
difference $\Delta T$ at $r=0.71R_\odot$, where  
$\Delta T =\max (T_1(r_\mathrm{bc},\theta))-\min
(T_1(r_\mathrm{bc},\theta))$. 
Although the temperature difference monotonously
increases with larger stellar angular velocity values, 
it is not enough to make the rotational profile largely deviate from the
Taylor-Proudman state.
This can be explained by using the thermal wind equation,
which is a steady state solution of 
eq. (\ref{therm01}):
\begin{eqnarray}
  0=
r\sin\theta\frac{\partial \Omega^2}{\partial z}
-\frac{g}{\gamma r}\frac{\partial s_1}{\partial \theta}.\label{therm02}
\end{eqnarray}
The inertial term and the diffusion term are neglected here.
This equation indicates that, for a given value of the NTP, we need an
entropy gradient proportional to $\Omega_0^2$. However, our simulation
results show that $\Delta T \propto \Omega_0^{0.58}$, which means that as
$\Omega_0$ increases, the thermal driving force becomes insufficient to
push differential rotation away from the Taylor-Proudman state.
In other words, the latitudinal entropy
gradient in rapidly rotating stars
is so small that differential rotation stays close to the
Taylor-Proudman
state. In our model, meridional flow
generates latitudinal entropy gradient at the base of the convection
zone.
It is conjectured that the insufficient thermal drive is due to
a slow meridional flow.\par
\begin{figure}[htbp]
 \epsscale{1.}
 \plotone{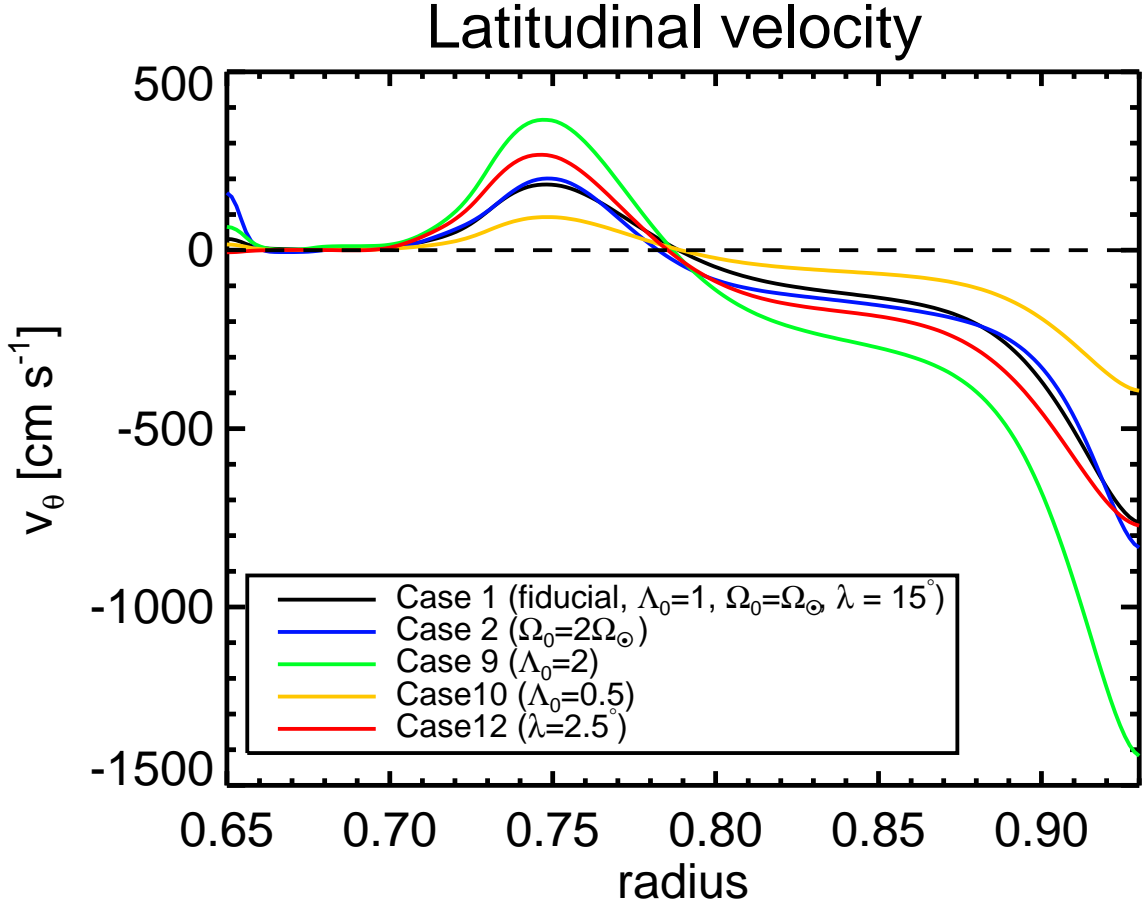}
 \caption{Profiles of latitudinal velocity ($v_\theta$) at
 colatitude $\theta=45^\circ$ as a function of radial distance.
 In case 1, stellar angular velocity is the solar value, and the
 amplitude of angular momentum transport $\Lambda_0=1$. In case 2,
 stellar angular velocity $\Omega_0=2\Omega_\odot$. In case 9 and 10,
 amplitude of the turbulent angular momentum transport  $\Lambda_0=2$
 and $0.5$, respectively.
 In case 12, the inclination angle of $\Lambda$ effect $\lambda=2.5^\circ$, and other
 parameters are the same as case 1.
\label{vari_some}}
\end{figure}

We next investigate the dependence of meridional
flow on stellar angular velocity.
Fig. \ref{vari_some} shows
the radial profile of latitudinal velocity $v_\theta$ at
$\theta=45^\circ$, using the results of cases
1, 2 and 9. In case 2, stellar angular velocity is twice that of case
1 (the solar value).
In case 9, stellar angular velocity is equal to the solar
value, and the amplitude of the $\Lambda$ effect is two times the
value in case 1. 
Fig. \ref{vari_some} shows that meridional flow does not depend on
stellar angular velocity, while it correlates with the $\Lambda$ effect.
Considering
eq. (\ref{lambda}), the $\Lambda$ effect increases 
with larger values of stellar angular velocity, since the amplitude of the
$\Lambda$
effect is proportional to $\Omega_0$.
The reason why differential rotation in rapidly rotation stars
is close to  the Taylor-Proudman state is that meridional flow does not
become
fast with large stellar angular velocity values.
\par
We interpret the result that the speed of meridional flow does not depend on
stellar angular velocity in our model as follows.
With large values of stellar angular velocity, more angular
momentum is transported by the $\Lambda$-effect (Note that the
$\Lambda$-effect is proportional to 
$\Omega_0$ in equation (\ref{amp_lambda})), so meridional flow obtains
more energy from differential rotation. The energy gain does not result
in an increase in speed because of the associated enhancement of the Coriolis
force, which bends the meridional flow in the longitudinal direction.
Another explanation is possible in terms of angular momentum
transport.
The angular momentum
fluxes from both meridional flow and the Reynolds stress ($\Lambda$ effect)
must be
balanced in a steady state. The former is proportional to
$v_\mathrm{m}\Omega_0$ and the
latter is proportional to $\Omega_0$, where $v_\mathrm{m}$ is the
amplitude of meridional flow. Therefore, meridional flow does not
depend on stellar angular velocity \citep{2005LRSP....2....1M}.
Our results (Fig. \ref{vari_some}) indicate
that with a larger stellar angular velocity (case 2),
the above mechanism does not generate fast 
meridional flow. However, this does not occur  when only the $\Lambda$
effect is large (case 9).
\subsection{Angular Velocity Difference on the
  Surface}\label{delta_omega}
In this subsection we discuss angular velocity difference
$\Delta \Omega$ at the surface and the relationship between our results
and previous observations. We conduct numerical simulations to investigate
the physical process which determines $\Delta \Omega$ (cases 1, 6-11). 
We define angular velocity difference as
$\Delta \Omega =
\max(\Omega_1(r_\mathrm{max},\theta))-\min(\Omega_1(r_\mathrm{max},\theta))$.
\par
$\Delta \Omega$ is determined
by two opposing effects, a smoothing effect from turbulent
viscosity and a steepening effect from the $\Lambda$ effect.
In a stationary
state these two effects cancel each other out. Latitudinal flux for
turbulent viscosity and the $\Lambda$ effect can be written as
$\rho_0\nu_\mathrm{0v}\Delta \Omega/\Delta \theta$
and $\rho_0\nu_\mathrm{0l}\Lambda_0\Omega_0$, respectively.
Because these two
have approximately the same value, $\Delta\Omega$ can be estimated as
\begin{eqnarray}
 \Delta \Omega
  \sim\frac{\nu_\mathrm{0l}}{\nu_\mathrm{0v}}\Lambda_0\Omega_0\Delta\theta,
\label{deltaomega}
\end{eqnarray}
where $\Delta \theta$ denotes the differential
rotation region.\par
\begin{figure}[htbp]
 \epsscale{1.}
 \plotone{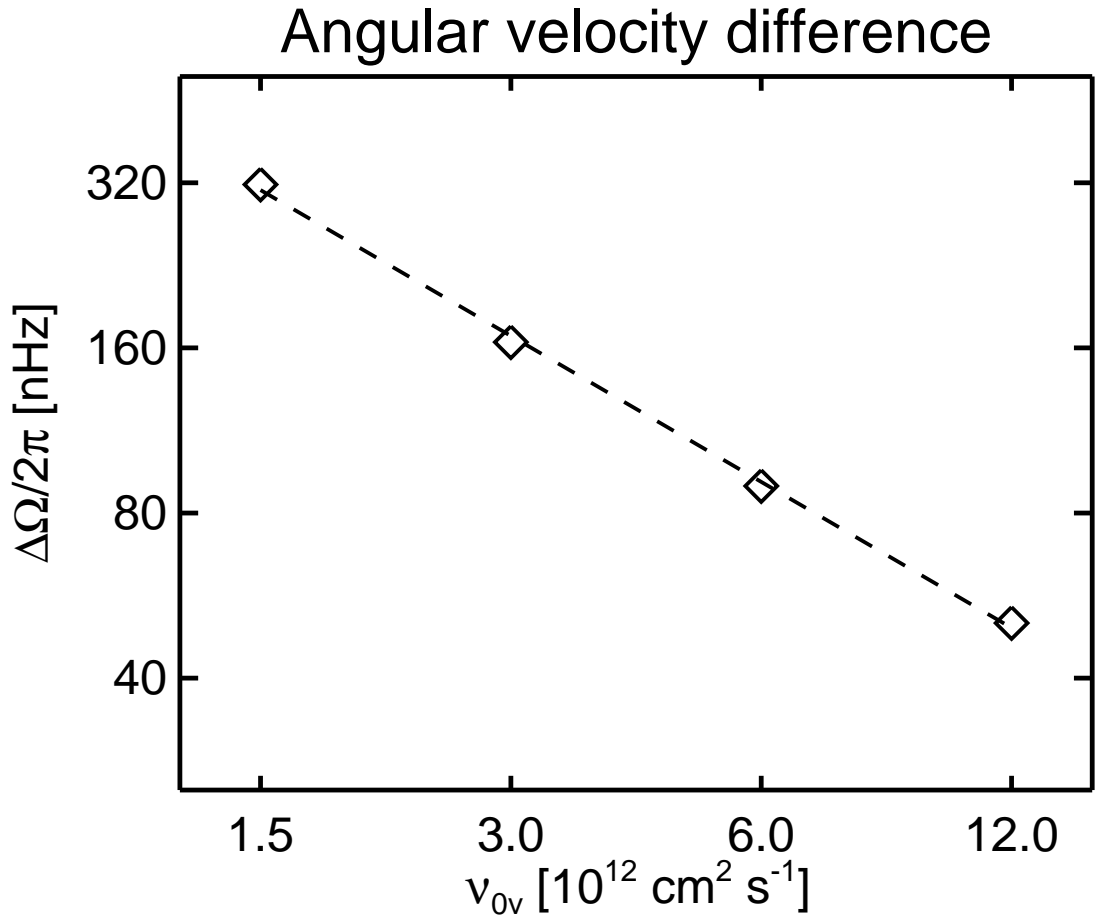}
 \caption{Angular velocity difference at the
 surface as a function of the coefficient of turbulent viscosity
 $\nu_\mathrm{0v}$. 
The dashed line is the fit to the results showing a power-law function
 with an index of $-0.88$.\label{vis_vari}}
\end{figure}
\begin{figure}[htbp]
 \epsscale{1.}
 \plotone{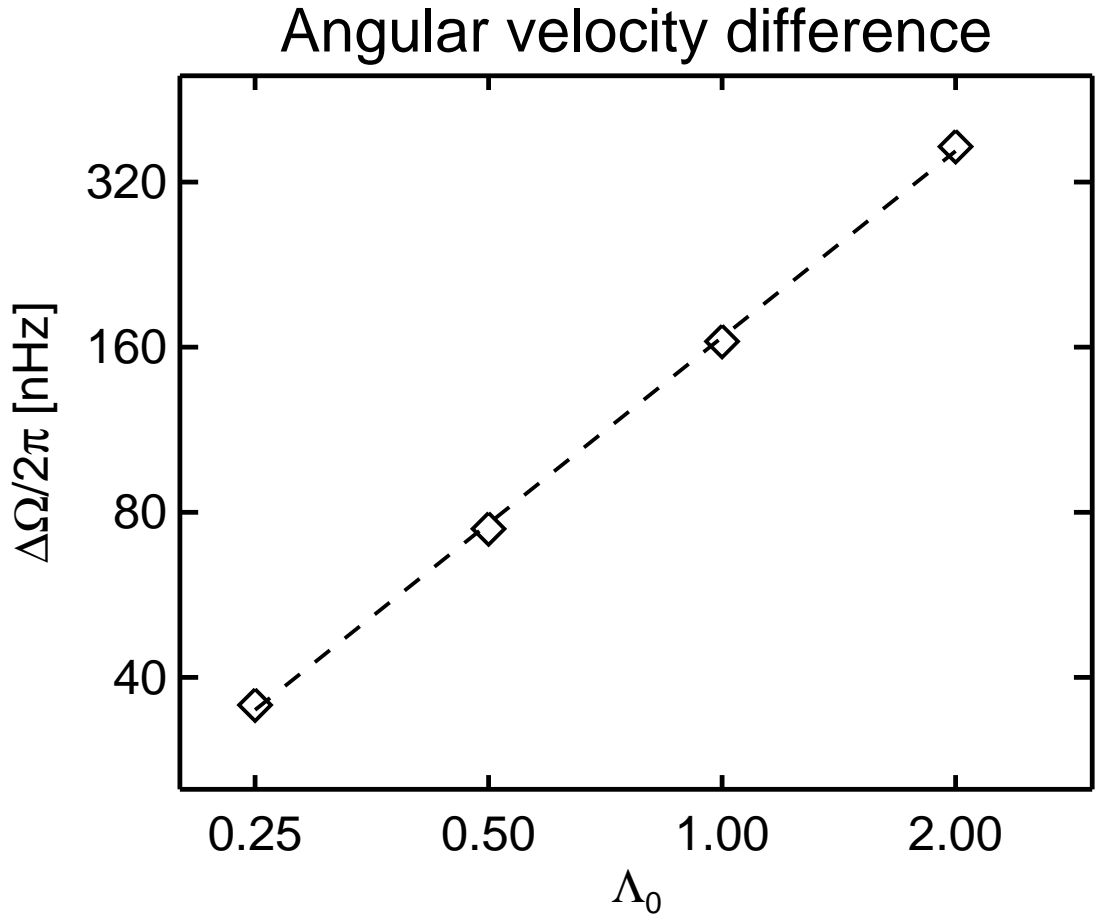}
 \caption{Angular velocity difference at the
 surface as a function of the amplitude of the angular momentum
 transport $\Lambda_0$. 
The dashed line is the fit to the results showing a power-law function
 with an index of $1.1$.\label{lam_vari}}
\end{figure}

In order to confirm eq. (\ref{deltaomega}),
we conduct two sets of simulations, firstly varying the value of
turbulent viscosity ($\nu_\mathrm{0v}$), and secondly the
amplitude of the
$\Lambda$ effect ($\Lambda_0$).
Note that the setting for turbulent viscosity does not reflect a real situation,
since the coefficients
of turbulent viscosity and the $\Lambda$ effect should have a 
common value.
Nonetheless, this is necessary for the purpose of our investigation.
The simulation results are shown in Figures \ref{vis_vari} and
\ref{lam_vari}. We obtain
$\Delta \Omega \propto \nu_\mathrm{0v}^{-0.88}$ and
$\Delta \Omega \propto \Lambda_0^{1.1}$, which
are consistent with eq. (\ref{deltaomega}).
\par
\begin{figure}[htbp]
 \epsscale{1.}
 \plotone{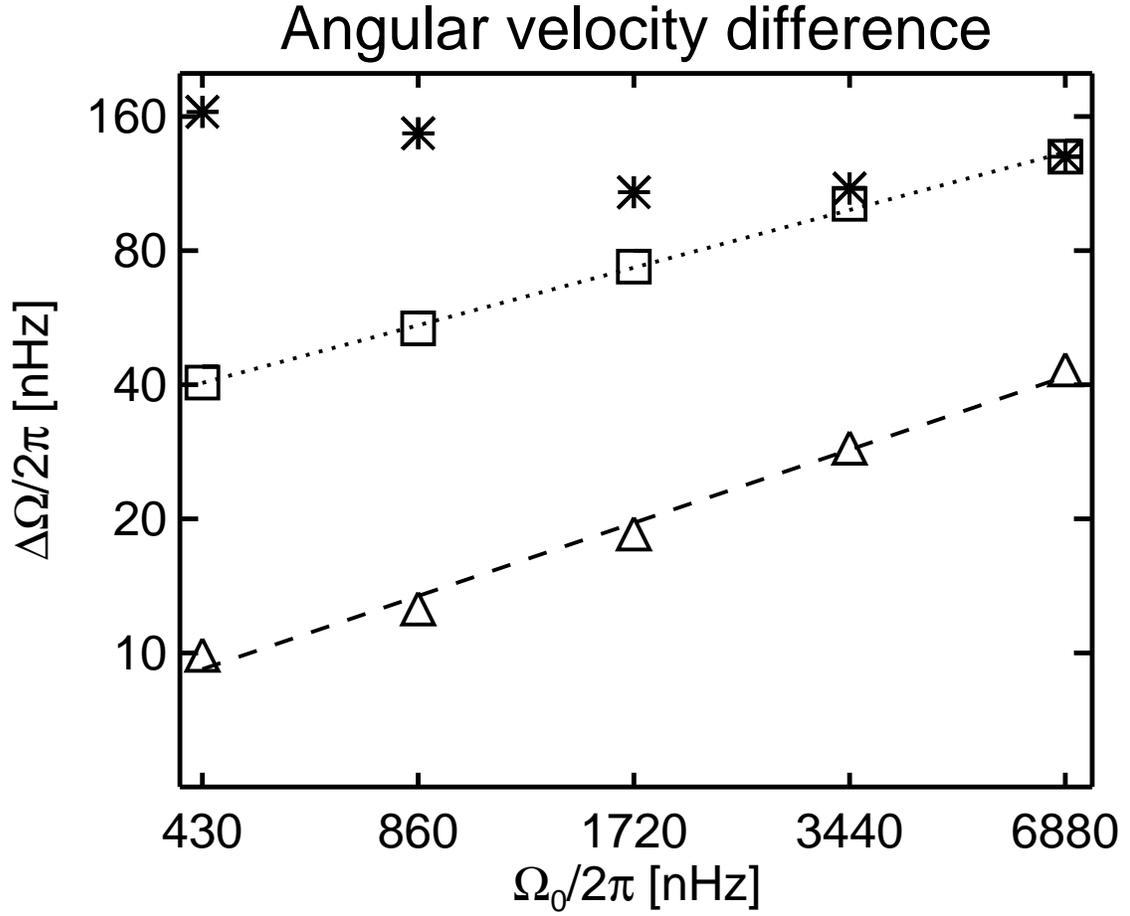}
 \caption{Angular velocity difference in three
 regions. Asterisks, squares and triangles represent the equator and the
 pole, the equator and the colatitude $\theta=45^\circ$ and the
 equator and the colatitude $\theta=30^\circ$, respectively.
The dashed and dotted lines are the fits to the results showing a power-law function
 with indices of $0.43$ (squares) and $0.55$ (triangles).
\label{angular_difference}}
\end{figure}

Fig. \ref{angular_difference} shows the results of the dependence of 
$\Delta \Omega$ on $\Omega_0$ (Cases 1-5).
Asterisks denote the difference at the surface between the
equator and the pole,
squares show the difference between the equator and the
colatitude $\theta=45^\circ$, and triangles are the difference between
the equator and the colatitude $\theta=60^\circ$.
The difference at low latitudes  (squares and triangles) monotonically
increases with stellar angular velocity.
However this is not the case for angular velocity difference between the
equator and the pole (asterisk).
As we
discussed in \S \ref{taylor}, 
when stellar rotation velocity is large, the Taylor-Proudman state is
achieved, meaning 
the gradient of angular velocity at the surface concentrates in
lower latitudes.
Due to this concentration, $\Delta\theta$ becomes
smaller in Eq. (\ref{deltaomega}) with larger values of $\Omega_0$. Thus,
$\Delta \Omega_0$ does not show an explicit dependence on $\Omega_0$.
At low
latitudes, $\Delta\theta$ is fixed and the angular velocity difference
increases with stellar angular velocity. We obtain
$\Delta\Omega\propto\Omega_0^{0.43}$ (between the equator and the colatitude
$\theta=45^\circ$: squares) and 
$\Delta\Omega\propto\Omega_0^{0.55}$ (between the equator and the colatitude 
$\theta=60^\circ$: triangles). 
This indicates that
$\Delta\Omega/\Omega_0 $ decreases with stellar angular velocity. These
results are consistent with previous stellar observations
\citep{1996ApJ...466..384D,2003A&A...398..647R,2005MNRAS.357L...1B}.
\subsection{Variation of $\Lambda$-effect and
  superadiabaticity}\label{variation}
In this section, we discuss the dependence of meridional flow and
differential rotation on free parameters. The parameter set is shown
in Table \ref{param} (cases 12-17). At first we investigate the
influence of the variation of the $\Lambda$ effect. The $\Lambda$ effect has two free
parameters, i.e., amplitude $\Lambda_0$ and inclination angle
$\lambda$ (see \S \ref{s:lambda}). Amplitude is thought to become
smaller with a larger stellar
angular velocity, 
due to the saturation of the correlations such as
$\langle v'_rv'_\phi \rangle$ and $\langle v'_\theta v'_\phi \rangle$, where
$v_r'$, $v_\theta'$ and $v_\phi'$ are the radial, latitudinal and
longitudinal component turbulent velocities, respectively.
Fig. \ref{vari_some} shows that
meridional flow becomes slower with a smaller $\Lambda_0$, keeping the
$\Omega_0$ value constant (Case 10). It is clear with the result of \S
\ref{taylor} that meridional flow becomes slow with a larger angular
velocity when the variation of
$\Lambda_0$ is included. \cite{2008ApJ...689.1354B} reported
this effect with their three-dimensional hydrodynamic calculation.
When meridional flow is
slow, the entropy gradient generated by the subadiabatic layer is
small, and differential rotation approaches the Taylor-Proudman state.\par
The inclination angle is thought to be small with large stellar angular
velocity values, since the motion across the rotational axis is restricted \citep{1993A&A...276...96K}. In
case 12, differential rotation with a small inclination angle
($\lambda=2.5^\circ$) is
calculated. Other parameters are the same as case 1. The radial
distribution of meridional flow is shown in Fig. \ref{vari_some}.
Meridional flow becomes faster with a smaller inclination angle.
Because of the efficient angular momentum transport in the $z$ direction
when the inclination angle is small, the second term on the right hand side
of Eq. (\ref{therm01}) is large. This generates a large
$\omega_\phi$, i.e. fast meridional flow. 
\par
In summary, we found that rapid stellar rotation
causes two opposing effects on the speed of meridional flow. The
speed is reduced by the suppression of $\Lambda_0$, while it is enhanced by the
angular momentum transport along the axial direction with a smaller
$\lambda$. Although the results of the three-dimensional calculation suggest that
meridional flow becomes slower with a larger stellar angular
velocity, our model cannot draw a conclusion about the speed of
meridional flow in rapidly rotating stars.
\par
\begin{figure}[htbp]
 \epsscale{1.}
 \plotone{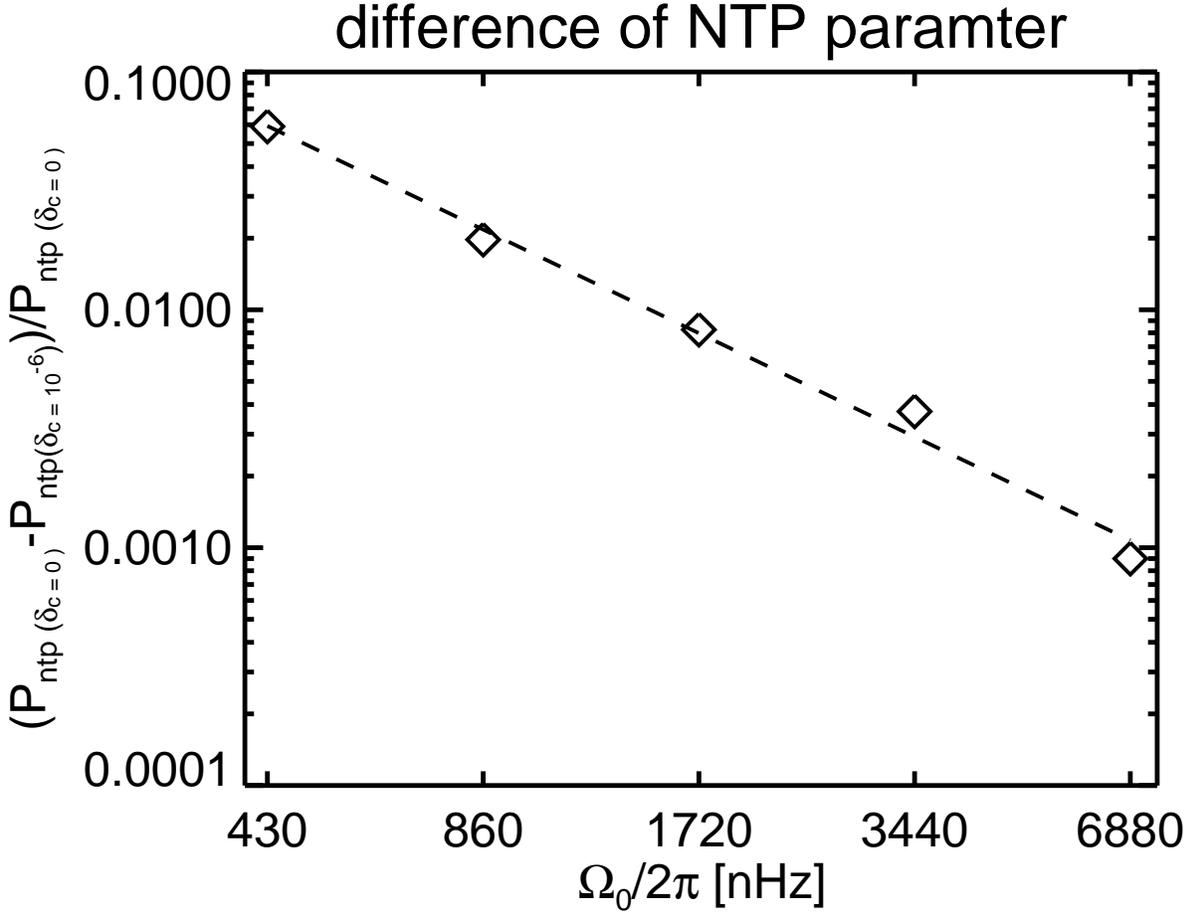}
 \caption{
The difference of the NTP paramters between cases with adiabatic and
 superadiabatic convection zone
 $(P_{\mathrm{ntp}(\delta_\mathrm{c}=0)}-P_{\mathrm{ntp}(\delta_\mathrm{c}=10^{-6})})/P_{\mathrm{ntp}(\delta_\mathrm{c}=0)}$
 as a
 function of stellar angular velocity $\Omega_0/2\pi$.
 The dashed line is the fit to the results showing a power-law function
 with an index of $-1.4$.\label{super}}
\end{figure}

Next we investigate the influence of superadiabaticity in the
convection zone. In cases 13-17, superadiabaticity in the convection
zone $\delta_\mathrm{c} = 1\times 10^{-6}$. The differences of the NTP
parameters with adiabatic and superadiabatic convection
zones  $(P_{\mathrm{ntp}(\delta_\mathrm{c}=0)}-P_{\mathrm{ntp}(\delta_\mathrm{c}=10^{-6})})/P_{\mathrm{ntp}(\delta_\mathrm{c}=0)}$
are shown in Fig. \ref{super}.
The NTP parameter values with a superadiabatic convection zone
are smaller than those with an adiabatic convection
zone, since meridional flow in the superadiabatic
convection zone makes the entropy gradient small.  This result is
suggested by \cite{2005ApJ...631.1286R}. Note that the difference
between the values of the NTP parameters with an adiabatic and those
with a
superadiabatic convection zone decreases as the stellar angular
velocity increases, since the generation of entropy gradient
by the subadiabtic layer becomes ineffective with a larger stellar
angular velocity.
\section{SUMMARY}
We have investigated differential rotation in rapidly rotating
stars using a mean field model.
This work is significant because it can be used as
a base for further research on
stellar activity cycles, which are most likely caused by the dynamo
action of differential rotation in the stellar convection zone.
\par
First, we investigated the
morphology of differential rotation in rapidly rotating stars.
Although more angular momentum is transported by
convection with larger stellar angular velocity, the Coriolis force is
stronger than in the solar case, so
meridional flow does not be fast.
In our model, meridional flow
generates latitudinal entropy gradient in the subadiabatically
stratified overshoot region. Since the meridional flow is not fast, the
entropy gradient is insufficient to
move differential rotation far from the Taylor-Proudman state in rapidly rotating stars.
As a result, the differential rotation of stars with large
stellar angular velocity is close to the Taylor-Proudman state.\par
The temperature difference between latitudes is probably controlled by
two important factors, i.e., the subadiabatic layer below
the convection zone and anisotropic heat transport caused by turbulence and
rotation. We
suggest that the former is important in slow rotators like the sun, and the
latter in rapid rotators.
The subadiabatic-layer effect is included in our model, while
anisotropic heat transport is not. 
We found that the effect of the subadiabatic layer can generate a temperature
difference $\Delta T=10\ \mathrm{K}$ in the solar case, which moderately
increases with higher rotation speeds, and
$\Delta T=30\ \mathrm{K}$ in case $\Omega_0=8\Omega_\odot$.
The three-dimensional simulations by \cite{2008ApJ...689.1354B} include
a self-consistent calculation of
anisotropy of turbulent thermal transport but not the
subadiabatic layer at the bottom boundary. In their calculation 
$\Delta T$ is most likely smaller than $10\ \mathrm{K}$ in the solar case, since they cannot
reproduce the solar differential rotation only with anisotropy of
thermal transport. Also, $\Delta T=100\ \mathrm{K}$ in case
$\Omega_0=5\Omega_\odot$, which is larger than the case with the subadiabatic layer. We speculate that anisotropic
heat transport becomes more significant in rapidly rotating stars.
There is also a possibility that our calculated entropy gradient at the base of
the convection zone can be used as a boundary condition for a
self-consistent three dimensional simulation of stellar
convection \citep{2006ApJ...641..618M}.
Note that differential rotation in rapidly rotating stars in
\cite{2001A&A...366..668K} is not in the Taylor-Proudman state when
anisotropy of turbulent thermal conductivity is included.
A future study of the simultaneous effects of the attached subadiabatic
layer beneath convection zone and anisotoropy of the turbulent thermal
conductivity on stellar differential rotation would provide a better
understanding of stellar differential rotation.
\par
Next, we investigated angular velocity difference at the surface.
The $\Lambda$ effect causes spatial difference in the rotation
profile, while turbulent viscosity reduces the difference.
Angular velocity difference $\Delta\Omega$ is determined in
eq. (\ref{deltaomega}), which is then used to investigate differential
rotation in rapidly rotating stars.
Since stellar rotation is close to the Taylor-Proudman state, and the
radiative core is rotating rigidly,
differential rotation is concentrated at low latitudes
with large stellar angular velocity.
This concentration leads to a small
$\Delta\theta$ in eq. (\ref{deltaomega}). 
Therefore, only at low latitudes our model
is consistent with stellar observations.
\par
 Our conclusions are as follows: (1) Differential rotation approaches
 the Taylor-Proudman state when stellar rotation is faster than
solar rotation.
(2) Entropy gradient generated by the attached subadiabatic layer
beneath the convection zone becomes relatively small with a large
stellar angular velocity.
(3) Turbulent viscosity and turbulent angular momentum
 transport determine the spatial difference of angular velocity
 $\Delta\Omega$. (4) The results of our mean field model can explain
 observations of stellar differential rotation.\par
Our future work will focus on the stellar MHD dynamo. 
Several investigations have been conducted on the stellar dynamo using a 
kinematic dynamo framework
\citep{2001ASPC..248..235D,2001ASPC..248..189C,2009A&A...497..829M,2010A&A...509A..32J}.
Since, under such a framework,
only the magnetic induction equation is solved using a given velocity
field, solving a linear equation,
such analysis does not give sufficient information on the strength
of the dynamo-generated stellar magnetic
field. 
To obtain the full amplitude of the stellar magnetic field,
the feedback
to the velocity field is required, i.e., an MHD framework.
Adopting a similar approach to \cite{2006ApJ...647..662R}, we can use the
results of this paper to investigate the strength of the stellar
magnetic field.
Recent observations of the strength of the magnetic
field generated by stellar differential rotation
have been conducted using spectroscopy
\citep[e.g.][]{2008MNRAS.388...80P}. A comparison of these observations
and numerical calculations of the stellar dynamo could give new insight into
the stellar magnetic field. 
Finally, our stellar MHD dynamo study would also contribute to the
understanding of recent investigations into  stellar magnetic cyclic
activity periods
\citep{1984ApJ...287..769N,1999ApJ...524..295S}.

\acknowledgements
We are most grateful to Dr. M. Rempel for helpful advice.
Numerical computations were carried out at the General-Purpose PC farm
in the Center for Computational Astrophysics (CfCA) of the National
Astronomical Observatory of Japan.
The page charge for this paper is supported by CfCA.
We have
greatly benefited from the proofreading/editing assistance from
the GCOE program.


\clearpage

\begin{table}
\begin{center}
\caption{Significant parameters of the simplified model.\label{param}}
\begin{tabular}{lccccc}
\tableline\tableline
Case & $\Omega_0\ \mathrm{[nHz]}$& 
$\nu_\mathrm{0v}\ \mathrm{[cm^2\ s^{-1}]}$& $\Lambda_0$ & $\lambda$ & $\delta_\mathrm{c}$\\
\tableline
1& $1\Omega_\odot=430$  & $3\times10^{12}$ & 1 &$15^\circ$ & 0\\
2& $2\Omega_\odot=860$  & $3\times10^{12}$ & 1 &$15^\circ$ & 0\\
3& $4\Omega_\odot=1720$ & $3\times10^{12}$ & 1 &$15^\circ$ & 0\\
4& $8\Omega_\odot=3440$ & $3\times10^{12}$ & 1 &$15^\circ$ & 0\\
5& $16\Omega_\odot=6880$& $3\times10^{12}$ & 1 &$15^\circ$ & 0\\
6& $1\Omega_\odot=430$  & $12\times10^{12}$ & 1 &$15^\circ$ & 0\\
7& $1\Omega_\odot=430$  & $ 6\times10^{12}$ & 1 &$15^\circ$ & 0\\
8& $1\Omega_\odot=430$  & $1.5\times10^{12}$ & 1 &$15^\circ$ & 0\\
9& $1\Omega_\odot=430$  & $3\times10^{12}$ & 2 &$15^\circ$ & 0\\
10& $1\Omega_\odot=430$  & $3\times10^{12}$ & 0.5 &$15^\circ$ & 0\\
11& $1\Omega_\odot=430$  & $3\times10^{12}$ & 0.25 &$15^\circ$ & 0\\
12& $1\Omega_\odot=430$  &  $3\times10^{12}$ & 1&$2.5^\circ$ & 0 \\
13& $1\Omega_\odot=430$  & $3\times10^{12}$ & 1 &$15^\circ$ & $1\times10^{-6}$ \\
14& $2\Omega_\odot=860$  & $3\times10^{12}$ & 1 &$15^\circ$ & $1\times10^{-6}$ \\
15& $4\Omega_\odot=1720$ & $3\times10^{12}$ & 1 &$15^\circ$ & $1\times10^{-6}$ \\
16& $8\Omega_\odot=3440$ & $3\times10^{12}$ & 1 &$15^\circ$ & $1\times10^{-6}$ \\
17& $16\Omega_\odot=6880$ & $3\times10^{12}$ & 1 &$15^\circ$ & $1\times10^{-6}$ \\
\tableline
\end{tabular}
\end{center}
\end{table}

\end{document}